\documentclass[aps,showpacs,floatfix,twocolumn,superscriptaddress]{revtex4-1}
\usepackage{graphicx}
\usepackage{amsfonts}
\usepackage{amssymb}
\usepackage{amsmath}
\usepackage{natbib}
\usepackage{dcolumn}
\usepackage{bm}
\begin{document}
\title {Nuclear equation of state in a relativistic 
independent quark model with chiral symmetry and variation with quark masses}
\author{N. Barik}
\affiliation{Department of Physics, Utkal University, Bhubaneswar-751 004, 
India}
\author{R.N. Mishra}
\author{D.K. Mohanty}
\affiliation{Department of Physics, Ravenshaw University, Cuttack-753 003, 
India}
\author{P.K. Panda}
\affiliation{Department of Physics, C.V. Raman College of Engineering, 
Bhubaneswar-752 054, India}
\author{T. Frederico}
\affiliation{Instituto Tecn\'ologico de Aeron\'atica, DCTA
12228-900 S\~ao Jos\'e dos Campos, SP, Brazil}
\begin{abstract}
We have calculated the properties of nuclear matter in
a self-consistent manner with quark-meson coupling mechanism
incorporating structure of nucleons in vacuum through
a relativistic potential  model; where the dominant confining interaction
for the free independent quarks inside a nucleon, is represented by a
phenomenologically average potential in equally mixed scalar-vector
harmonic form. Corrections due to spurious centre of mass motion
as well as those due to other residual interactions such as
the one gluon exchange at short distances and quark-pion coupling arising
out of chiral symmetry restoration; have been considered in a perturbative
manner to obtain the nucleon mass in vacuum. The nucleon-nucleon interaction
in nuclear matter is then realized by introducing additional
quark couplings to sigma and omega mesons through mean field approximations.
The relevant parameters of the interaction are obtained
self consistently while realizing the saturation properties such as
the binding energy, pressure and compressibility of the nuclear matter.
We also discuss some implications of chiral symmetry in nuclear matter along with 
the nucleon and nuclear sigma term and the sensitivity of nuclear matter
binding energy with variations in the light quark mass.
\end{abstract}

\maketitle
\section{Introduction}
The properties of nuclear matter has been an area of considerable interest
for the past few decades. Such studies are quite important in nuclear
physics, (e.g. in the context of nucleon-nucleon (N-N) interaction, structure
and properties of finite nuclei, and dynamics of  heavy ion collisions),
astrophysics (nucleosynthesis, structure and evolution of neutron stars
\cite{prakash}, big bang cosmology) and also particle physics (production
or interaction of hadrons).  One of the fundamental concerns in the study
of nuclear matter is the nature of the N-N interaction. This problem is solved
usually in a self-consistent manner in various different approaches
which can be broadly classified into three general types, namely the 
{\it ab initio} methods, the effective field theory method and
methods based on phenomenological density functionals.
The {\it ab initio} methods include the Brueckner-Hartree-Fock (BHF) approach
\cite{jamion,zhou,baldo}; the relativistic Dirac-Brueckner-Hartree-Fock
(DBHF) approach \cite{brockmann,li,jong,dalen}, the Green Function
Monte-Carlo (GFMC) method \cite{carlson,dickhoff,fabrocini} using the
basic N-N interactions given by boson exchange potentials. The other
approach of this  type pioneered by the Argonne Group \cite{akmal97,akmal98}
is also known as the  variational approach.
The effective field theory (EFT) methods \cite{furnstahl} are based on
chiral perturbation theory \cite{lutz,finelli}.
The third type of approaches are based on
the phenomenological models with effective density dependent interactions
such as Gogny or Skyrme forces \cite{bender} (see also  \cite{dutra}
for a systematic analysis of Skyrme models)  and the relativistic mean
field (RMF) models \cite{qhd}.
The parameters of these models are evaluated by appealing to the bulk
properties of nuclear matter and properties of closed shell nuclei.

The RMF-models represent the N-N interactions through the coupling of
nucleons with isoscalar scalar mesons, isoscalar vector mesons, isovector
vector mesons and the photon quanta besides the self and cross-interactions
among these mesons \cite{nl3,horowitz-01}. Although implemented at Hartree
level only, these models have been very successful in simulating the observed
bulk properties of nuclear matter including the nuclear equation of state
(EOS), mass and radii of neutron star as well as in explaining properties
of finite nuclei. Recently, the RMF model has also been extended to include the
Hartree-Fock theory and the short range repulsion using unitary operator method
\cite{short} to study the symmetric nuclear matter.

In all these approaches mentioned above nucleons are treated as structureless 
point objects. However, incorporation of stucture of nucleon with meson 
couplings at the basic quark level in the study of saturation properties of
nuclear matter can provide new insight. With such a hope there has been
several attempts based on simple bag-model or some phenomenological potential 
models to address the nucleon structure. Using such quark-meson coupling (QMC)
models nuclear equations of state (EOS) have also been constructed
\cite{guichon, frederico89} and properties of nuclear matter have been 
studied in great detail in a series of works by Saito, Thomas
and collaborators \cite{ST} and by others \cite{recent,temp,phase,qmcparm}.

Quantum Chromodynamics (QCD) is the underlying theory of the strong forces
that hopefully would also explain nuclear stability. Therefore the study
of changes in nuclear properties due to the fundamental parameters of QCD
in particular  with the variation of light quark masses is legitimate.
The sensitivity of the relative binding energy to the relative change in
light quark masses was investigated for nuclei A=3-8 and computed in
\cite{flambaum07}  for different Argonne potential models, considering cases
including the Urbana model IX three-body force, and for thorium in
\cite{flambaum09}. More recently, Ref. \cite{carillo} computed the variations
in the nuclear binding of light elements like deuteron,
tritium, $^7$Li, $^{12}$C and $^{16}$O from the changes in quark masses
in a one boson-exchange model. Although, there are sizable
differences between \cite{flambaum07} and \cite{carillo},
they agree within a factor less than 2, which can be due to different
nuclear models, and the detailed form of computing the meson mass and
coupling constant variations.  In another study, the anthropomorphic principle
was used to constrain variations in the quark masses and fine structure
constant from the abundances of carbon and oxygen in the universe essential
for life \cite{epelbaum}. The nuclear equations of state and the nuclear
matter binding energy, can also be affected by quark mass variations with
interesting implications for the stability of heavy nuclei and stars. This
motivates a study of such effect within relativistic models of nuclear matter
such as  QMC models.

In the present study, we have developed a modified quark-meson coupling
model (MQMC) \cite{barik,frederico89,batista}, which is based on a suitable
relativistic independent quark potential model rather than a bag to address the 
nucleon structure in vacuum. In such  a picture the light quarks
inside a bare nucleon are considered to be independently confined
by a phenomemenologically average potential with equally mixed scalar-vector
harmonic form. Such a potential has characteristic simplifying features in
converting the independent quark Dirac equation into an effective
Schr\"odinger like equation for the upper component of the Dirac spinor which 
can be easily solved. The implication of such potential forms in the Dirac
framework has been studied earlier by several authors \cite{smith}.
It has been shown that the spin-orbit interaction is absent in such models
due to exact cancellation of terms coming from the vector and the scalar
part of the potential taken in equal proportonal. This is a welcome feature
for baryon sector, where the contribution from spin-orbit interaction to
baryon mass spliitings is already known to be negligible \cite{feynmann}.

Eichen and Feinberg \cite{eichen} in a gauge invariant formalism, assuming the
confinement mechanism to be purely color-electric in character, obtained 
similar Lorentz structure of the potential. This typical Lorentz structure of 
the confining potential renders Dirac equation solvable for all possible quark
eigen-modes. Due to the harmonic nature of the potential; the quark orbitals
corresponding to the lowest eigen-mode is realised here in the familiar 
Gaussian form that makes the perturbative treatment of the residual 
interactions such as the short range one-gluon exchange and quark pion 
coupling arising out of chiral symmetry restoration in PCAC-limit as well 
as that due to the spurious centre of mass motion in the ground state, 
simple and straight forward in
comparison with other models. Therefore, it has provided a very suitable
alternative to the otherwise successful cloudy bag-models and has been 
extensively applied with remarkable consistency in baryonic as well as mesonic
sector\cite{appelquist, bdd}. 

Taking gluonic and pionic corrections
together with that due to centre of mass motion; baryon mass spectra in vacuum
had been successfully reproduced in this potential model \cite{frederico89}.
This model has also been quite successful in studying nucleon structure
functions in deep inelastic scattering \cite{dis}. In view of this we would 
like to adopt this model here to address the nucleon structure properties
of nucleons and nuclear matter.

Corrections due to the spurious centre of mass motion as well as those
due to short range one gluon exchange and quark -pion coupling would be
accounted for in a perturbative manner to obtain the nucleon mass in vacuum.
Then the (N-N) interaction in nuclear matter is realized by introducing
additional quark coupling to sigma ($\sigma$) and omega ($\omega$) mesons
through a mean field approximations. The relevant parameters of the interaction
 are obtained self consistently while realising the saturation properties
such as the binding energy, pressure and compressibility of the nuclear matter.
We examine the effective nucleon mass, nuclear sigma term as well as the
effective quark- condensate  at saturation density in comparison  with the
respective values at zero density. We also study their variations including
the sensitivity of the nuclear matter binding energy with the variation of
the light quark masses.

The paper is organized as follows: In Sec. II, we provide a brief outline
of the model describing the nucleon structure in vacuum where the nucleon mass 
can be obtained by appropriately taking into account the centre of mass correction, 
pionic correction and gluonic correction. The mean-field properties of
symmetric nuclear matter in this model is discussed in Sec. III. The 
results and discussions are presented in Sec. IV. Finally, in the last section, 
the conclusions are drawn.

\section{Potential model}

We choose from a phenomenological point of view a flavor independent potential 
$U(r)$ confining the constituent quarks inside the nucleon in
accordance with \cite{barik}, where $U(r)$ is 
\[
U(r)=\frac{1}{2}(1+\gamma^0)V(r)
\]
with 
\begin{equation}
V(r)=(ar^2+V_0),~~~~~ ~~~ a>0. 
\label{eq:1}
\end{equation}
Here $(a,~ V_0)$ are the potential parameters. This confining interaction is believed to
provide phenomenologically the zeroth order quark dynamics of the hadron, and 
corrections, like gluon exchange can be added perturbatively. 
The quark Lagrangian density corresponding to the confining model  
\begin{equation}
{\cal L}^{0}_{q}(x)={\bar \psi}_{q}(x)\;[\;\frac{i}{2}\gamma^{\mu}
\overleftrightarrow\partial_{\mu}-m_{q}-U(r)\;]\;\psi _{q}(x),
\label{eq:2}
\end{equation}
leads to the Dirac equation for an individual quark as
\begin{equation}
[\gamma^0\epsilon_q-{\vec \gamma}.{\vec p}-m_q-U(r)]\psi_q(\vec r)=0 \ .
\label{eq:3}
\end{equation}
The normalized quark wave function $\psi_q(\vec r)$ can be written in
the two component form for the ground state as
\begin{eqnarray}
\psi_q(\vec r)=\frac{1}{\sqrt{4\pi}}\left(
\begin{array}{c}
i~g(\vec r)/r\\
\vec\sigma\cdot\hat r ~f(\vec r)/r
\end{array}\;\right)\chi_s.
\label{eq:4}
\end{eqnarray}
Defining 
\begin{equation}
(\epsilon^{\prime}_q+m^{\prime}_q)=(\epsilon_q+m_q)\equiv\lambda_q,
\label{eq:8}
\end{equation}
with  
\begin{eqnarray}
\epsilon^{\prime}_q= (\epsilon_q-V_0/2),~~~~~~&& m^{\prime}_q=(m_q+V_0/2)
\nonumber\\
~~~\mbox{and} ~~~~&& r_{0q}=(a\lambda_q)^{-1/4},
\label{eq:9}
\end{eqnarray}
it can be shown that the upper and lower components of $\psi_q(r)$
corresponding to the quark-flavor $q$ for the ground state $1s_{1/2}$ are
\begin{eqnarray}
g_{q}(r)\;&=&\;{\cal N}_q\left(\frac{r}{r_{0q}}\right)\exp {(-r^2/{2r^2_{0q}})},
\nonumber\\
f_{q}(r)\;&=&\;-\frac{{\cal N}_q}{\lambda_qr_{0q}}\;\left(\frac{r}{r_{0q}}
\right)^{2}\; \exp {(-r^{2}/{2r^2_{0q}})},
\label{eq:10}
\end{eqnarray}
where the normalization ${\cal N}_q$, is given by
\begin{equation}
{\cal N}^{2}_{q}=\frac{8\lambda_q}{\sqrt {\pi}r_{0q}}~
\frac{1}{(3\epsilon^{\prime}_{q}+m^{\prime}_{q})}.
\end{equation}
In the above, $\epsilon_q$ is the ground state $1s_{1/2}$ individual 
quark energy obtained from the eigenvalue condition
\begin{equation}
(\epsilon^{\prime}_q-m^{\prime}_q)\sqrt \frac{\lambda_q}{a}=3.
\label{eq:11}
\end{equation}
The solution of equation (\ref{eq:11}) for the quark  energy 
$\epsilon_q$ immediately leads to the zeroth order energy of the nucleon
\begin{equation}
E_N^0=\sum_q~\epsilon_q
\label{eq:12}
\end{equation}
We can now construct the nucleon state $|N\rangle$ as the symmetrized 
product of the spin-flavor wavefunction of the three independent quarks 
each in its ground state as in equation (\ref{eq:4}).

Considering the quark confinement inside the nucleon through the
phenomenological interaction potential $U(r)$, the model expression for the
zeroth order energy $E_N^0$ of the nucleon core is obtained as in
equation (\ref{eq:12}). However, there
may be appropriate correction to $E_N^0$ due to possible residual interactions
such as the quark gluon interaction at short distances originating from one
gluon exchange and quark pion interaction arising out of the requirement for
restoration of chiral symmetry at the $SU(2)\times SU(2)$ level
as well as that coming from the spurious center of mass motion of the ground
state nucleon. We next consider these corrections for the zeroth order energy 
$E_N^0$ of the nucleon core as follows.

\subsection{Center of mass correction}
In this model, the quark constituents are independently bound by a 
potential with fixed center to obtain the quark orbitals in the nucleon, 
which are used to construct a composite nucleon wave function. If the 
composite nucleon is to be considered as a translationally invariant state, its 
wave function must be corrected for the effects of spurious center of mass (cm) motion.
For center of mass correction, earlier workers \cite{barik} had 
followed the prescription given by Peierls-Yoccoz \cite{cm}. 
However, here we will extract the center-of-mass energy to 
first order in the difference between the fixed center and relative quark 
co-ordinates, using the method described by Guichon {\it et al}
\cite{guichon}. 

We assume that the Hamiltonian, $H_N$ for the composite nucleon 
can be written as 
\begin{equation}
H_N=H_{in}+H_{cm}
\label{eq:13}
\end{equation}
where $H_{in}$ is the Hamiltonian corresponding to the internal degrees of 
freedom and $H_{cm}$ is the center-of-mass  Hamiltonian.
We can write the total Hamiltonian from the equation (\ref{eq:2}) as
\begin{equation}
H_N=\int{\cal H}_{N} ~d^3 x
\label{eq:13a}
\end{equation}
Thus, the Hamiltonian density can be written as
\begin{equation}
\hat{\cal H}_N={\sum_{i=1}^{3}} \gamma_0(i)\Big[\vec\gamma(i)\cdot 
{\vec p_i} +m_q +{\frac{1}{2}}(1+{\gamma}_0(i))U(r_{i})\Big]
\label{eq:14}
\end{equation}
The internal Hamiltonian density can be written in a similar way in terms of 
the relative rather than the fixed coordinates, as 
\begin{equation}
\hat{\cal H}_{in}= {\sum_{i=1}^{3}}\gamma_0(i)\Big[\vec\gamma(i)\cdot 
{\vec \pi_i} +m_q +{\frac{1}{2}}(1+{\gamma}_0(i))U(\rho_i)\Big]
\label{eq:15}
\end{equation}
where 
\begin{equation}
{\vec \pi_i}={\vec p_i}-{\frac{1}{3}}{\sum_{j=1}^{3}} {\vec p_j}~~ {\rm and}~~
{\vec \rho_{i}}={\vec r_i}-{\frac{1}{3}}{\sum_{j=1}^{3}}{\vec r_j}={\vec r_i}-
{\vec R_{cm}}.
\label{eq:16}
\end{equation}
The center of mass contribution to the Hamiltonian density is then the 
difference between the two:
\begin{eqnarray}
\hat{\cal H}_{cm}&=&\hat{\cal H}_{N}-\hat{\cal H}_{in}
= {\sum_{i=1}^{3}} \gamma_0(i)\Big[{\frac{1}{3}}{\vec 
\gamma}(i)\cdot {\sum_{j=1}^{3}}{\vec p_j}\nonumber\\ 
&+& {\frac{1}{2}}(1+{\gamma}_0(i))\left[U(r_i)-U(\rho_{i})\right ]\Big].
\label{eq:17}
\end{eqnarray}
We now estimate the center of mass contribution to the nucleon energy by 
calculating the expectation value of the center of mass Hamiltonian.
Here, we take the solutions for the quark orbitals as
given in equation (\ref{eq:10}), and the 
composite nucleon spin flavor configuration $|N\rangle$ as per $SU(6)$ 
prescription. Now, we have
\begin{eqnarray}
\epsilon_{cm}&=& \langle N|\hat{\cal H}_{cm}|N\rangle\nonumber\\
 &= & \langle N|\hat{\cal H}_{cm}^{(1)}|N\rangle + 
\langle N|\hat{\cal H}_{cm}^{(2)}|N\rangle
\end{eqnarray}
where 
\begin{eqnarray} 
\langle N|\hat{\cal H}_{cm}^{(1)}|N\rangle 
&=& {\frac{1}{3}}\langle N|
{\sum_{i=1}^{3}}{\gamma}_0(i){\vec \gamma}(i)\cdot{\sum_{j=1}^{3}} {\vec p_j}
|N\rangle \nonumber\\
&=& {\frac{1}{3}}\langle N|
{\sum_{i=1}^{3}}{\gamma}_0(i){\vec \gamma}(i)\cdot {\vec p_i}
|N\rangle \nonumber\\
&=& \frac{6}{(3\epsilon_u^{\prime}+m_u^{\prime})r_{0u}^2}. 
\end{eqnarray}
In the above expression the terms $j\ne i$ infact vanish effectively. 
\begin{eqnarray} 
&&\langle N|\hat{\cal H}_{cm}^{(2)}|N\rangle \nonumber\\ 
&=& {\frac{1}{2}}\langle N|
{\sum_{i=1}^{3}}(1+ {\gamma}_0(i))[U(r_i)-U(\rho_i)]
|N\rangle\nonumber\\
&=& {\frac{1}{2}}\langle N|{\sum_{i=1}^{3}}
(1+ {\gamma}_0(i))(2{\vec r_i}\cdot {\vec R_{cm}}- R_{cm}^2)|N\rangle\nonumber\\
&=&\frac{23\epsilon_u^{\prime}+13m_u^{\prime}}{3(3\epsilon_u^{\prime}+
m_u^{\prime})^2r_{0u}^2} \ .
\end{eqnarray}
Thus, the total center of mass correction comes out as
\begin{equation}
\epsilon_{cm}=\frac{(77\epsilon_u^{\prime}+31m_u^{\prime})}
{3(3\epsilon_u^{\prime}+m_u^{\prime})^2r_{0u}^2} \ .
\end{equation}

\subsection{Chiral symmetry and pionic corrections}

Under the global, infinitesimal chiral transformation, we have
\begin{equation}
\psi_q(x)\longrightarrow \psi_q(x) - i\frac{{\mathbf \tau}\cdot{\mathbf \alpha}}{2}
\gamma^{5}\psi_q(x) \ .
\end{equation}
Substituting the above expression in the zeroth order Lagrangian and 
with little algebra, we get
\begin{equation}
{\cal L}^{0'}_{q}(x)\longrightarrow {\cal L}^{0}_{q}+
i(m_q+V(r)){\bar \psi_q(x)}\gamma^{5}({\mathbf \tau}
\cdot{\mathbf \alpha})\psi_q(x).
\end{equation}
The axial vector current of the quarks is not conserved as the scalar term
proportional to $G(r)=(m_q+V(r)/2)$ in the Lagrangian density 
${\cal L}_q^0$ is chirally odd.
The vector part of the potential poses no problem in this respect.
Therefore, in the present model, chiral symmetry in $SU(2)-$ sector is
restored by introducing in the usual manner, an elementary pion field
${\mathbf \phi}$ of small but finite mass $m_{\pi}\simeq 140$ MeV through the
additional terms in the original Lagrangian density ${\cal L}_{q}(x)$,
so as to write,
\begin{equation}
{\cal L}_q(x)={\cal L}^{0}_{q}(x)+{\cal L}^{0}_{\pi}(x)+{\cal L}^{\pi}_{I}(x)
\end{equation}
where,
\begin{equation}
{\cal L}^{0}_\pi(x)=\frac{1}{2}(\partial_{\mu}{\mathbf \phi})^2-
\frac{1}{2}m_{\pi}^2{\mathbf \phi}^2.
\label{lpi}
\end{equation}
The Lagrangian density ${\cal L}^{\pi}_{I}(x)$ corresponding to the quark pion
interaction is taken to be linear in isovector pion field ${\mathbf \phi}$
such that
\begin{eqnarray}
{\cal L}^{\pi}_{I}(x)&=&
-\frac{i}{f_{\pi}}G(r)~{\bar \psi_q(x)}{\gamma}^5
({\mathbf \tau}\cdot {\mathbf \phi})\psi_q(x)\nonumber\\
&\equiv&-i~ G_{qq\pi}~{\bar \psi_q(x)}{\gamma}^5
({\mathbf \tau}\cdot {\mathbf \phi})\psi_q(x) \ ,
\label{intlpi}
\end{eqnarray}
where $f_{\pi}\simeq 93 MeV$ is the phenomenological pion decay constant
and $G_{qq\pi}$ is the effective quark-pion coupling strength.
Then, the four-divergence of the total axial vector current becomes 
$\partial_\mu A^\mu(x)=-f_\pi m_\pi^2 \phi(x)$ and gives the partial conserved
axial current (PCAC) relation. Now the Hamiltonian from equation (\ref{lpi})
in second quantized form is given by
\begin{equation}
H_\pi=\sum_j\int d^3k ~ w_k~ \hat a_j(\vec k)^\dagger a_j(\vec k) \ ,
\label{hpi}
\end{equation}
where $\hat a_j^\dagger(\vec k)$ and $\hat a_j(\vec k)$ are the pion 
creation and annihilation
operators and $w_k=(k^2+m_\pi^2)^{1/2}$ is the pion energy. Finally
the interaction Hamiltonian corresponding to ${\cal L}^{\pi}_{I}(x)$ becomes
\begin{eqnarray}
H^\pi_I= -\frac{1}{(2\pi)^{3/2}}\sum_{B,B',j}\int d^3k &\Big[& V^{BB'}_j(k) 
\hat b_{B'}^\dagger \hat b_B \hat a_j(\vec k)\nonumber\\
&+& h.c.\Big],
\label{inthpi}
\end{eqnarray}
where $j$ corresponds to the pion-isospin index and $h.c.$ denotes the
hermitian conjugate. In the above equation, $\hat b_B^\dagger$ and 
$\hat b_B$ are the creation and annihilation operators of the baryon 
state with quantum
numbers of $N,\Delta\cdots$ etc. $V^{BB'}_j(k)$ represents
the baryon pion absorption vertex function in the point-pion approximation 
and is obtained as \cite{barik}
\begin{eqnarray}
V^{BB'}_j(k)&=&-\frac{i}{f_{\pi}}\frac{1}{(2w_k)^{1/2}}\int d^3r ~G(r)~
e^{ik\cdot r}\nonumber\\
&\times&\langle B'|\bar \psi_q(r){\gamma}^5~\psi_q(r)\tau_j|B\rangle
\label{vbb}
\end{eqnarray}
Assuming all the quarks in the initial and final baryon are in a
$1s_{1/2}$ state, then equation (\ref{vbb}) becomes
\begin{eqnarray}
V^{BB'}_j(k)&=&\frac{i}{f_{\pi}}\frac{1}{(2w_k)^{1/2}}
\sqrt{\frac{\pi}{2}}\frac{{\cal N}_q^2}{k^{3/2}\lambda_qr_{0q}^4}\nonumber\\
&\times&\langle B'|\sum_q({\sigma}_q\cdot\vec k)\tau_j|B\rangle ~I(k) \ ,
\label{vbb1}
\end{eqnarray}
where
\begin{equation}
I(k)=2\int_0^\infty dr ~r^{5/2} ~G(r)~ J_{3/2}(kr)~ e^{-r^2/r_{0q}^2}.
\end{equation}
The coupling of the non-strange quarks to the pion causes a shift in the energy of the 
baryon core. From the second order perturbation theory, the pionic 
self-energy is given by 
\begin{equation}
{\Sigma}_B(E_B)=\sum_k\sum_{B^{'}}{\frac{V^{{\dagger}{BB^{'}}}V^{BB^{'}}}
{E_B-w_k-M_{B^{'}}^0}}    
\end{equation}
where $ \sum_k=\frac{1}{(2\pi)^3}\int d^3 k$
and $B^\prime$ is the intermediate baryon state. For degenerate intermediate 
states on the mass-shell with $ M_B^0=M_{B^\prime}^0,$ the self-energy 
correction becomes \cite{barik}
\begin{eqnarray}
\delta M_{B}^\pi&=&\sum_ {B}(E_B=M_B^0=M_{B^\prime}^0)\nonumber\\
&=&-\sum_{k,{B^\prime}}\frac{V^{{\dagger}{BB^\prime}}V^{BB^\prime}}{w_{k}}
\end{eqnarray}
Using the explicit expression (\ref{vbb1}) for $V^{BB^\prime}(k)$, one gets
\begin{equation}
\delta M_{B}^\pi=-\frac{1}{3}I_{\pi}\sum_{B^\prime}C_{BB^\prime}
f_{BB^\prime\pi}^2 \ ,
\end{equation}
where 
\[C_{BB'}=(\sigma^{BB'}\cdot\sigma^{BB'})(\tau^{BB'}\cdot\tau^{BB'}).\]
For intermediate baryon states $B'$ we consider only the octet and decouplet
ground states.  Now putting the values of $f_{BB^\prime\pi}$ and
$C_{BB^\prime}$, we get the pionic self energy for the nucleon\cite{barik} 
\begin{equation}
\delta M_{N}^\pi=- \frac{171}{25}I_{\pi}f_{NN\pi}^2,
\label{pion-corr}
\end{equation}
where 
\begin{equation}
I_{\pi}=\frac{1}{\pi{m_{\pi}}^2}\int_{0}^{\infty}dk. 
\frac{k^4u^2(k)}{w_k^2}
\end{equation} 
with the axial vector nucleon form factor given as 
\begin{equation}
u(k)=\Big[1-\frac{3}{2} \frac{k^2}{{\lambda}_q(5\epsilon_q^{\prime}+
7m_q^{\prime})}\Big]e^{-k^2r_0^2/4} \ .
\end{equation}
The psedovector nucleon pion coupling constant $f_{NN{\pi}}$ can be obtained 
from the familiar Goldberg Triemann relation using the axial vector coupling 
constant value $g_A$ in the model.

\subsection{Gluonic corrections}
The one-gluon exchange interaction is provided by the interaction Lagrangian
density
\begin{equation}
{\cal L}_I^g=\sum J^{\mu a}_i(x)A_\mu^a(x) \ ,
\end{equation}
where $A_\mu^a(x)$ are the octet of gluon vector-fields and $J^{\mu a}_i(x)$ is the
$i$-th quark color current. The gluonic correction can be separated in two 
pieces, namely, one from the color electric field ($E^a_i$) and another one 
from the magnetic one ($B^a_i$) 
generated by the $i$-th quark color current density
\begin{equation}
J^{\mu a}_i(x)=g_c\bar\psi_q(x)\gamma^\mu\lambda_i^a\psi_q(x) \ ,
\end{equation}
with $\lambda_i^a$ being the usual Gell-Mann $SU(3)$ matrices and
$\alpha_c=g_c^2/4\pi$. The contribution to the mass due to the relevant
diagrams can be written as a sum of a color electric and magnetic part as
\begin{equation}
(\Delta E_N)_g=(\Delta E_B)_g^E+(\Delta E_B)_g^M \ ,
\end{equation}
where
\begin{eqnarray}
(\Delta E_N)_g^E &=&\frac{1}{8\pi}\sum_{i,j}\sum_{a=1}^8\int\frac{d^3r_id^3r_j}
{|r_i-r_j|}\nonumber\\
&\times&\langle B |J^{0 a}_i(r_i)J^{0 a}_j(r_j)|B\rangle \ ,
\end{eqnarray}
and
\begin{eqnarray}
(\Delta E_N)_g^M&=&-\frac{1}{8\pi}\sum_{i,j}\sum_{a=1}^8\int\frac{d^3r_id^3r_j}
{|r_i-r_j|}\nonumber\\
&\times& \langle B |\vec J^a_i(r_i)\vec J^a_j(r_j)|B\rangle \ .
\end{eqnarray}
\begin{figure}[ht]
\centering
\vspace{-0.8in}
\includegraphics[width=6.cm,angle=0]{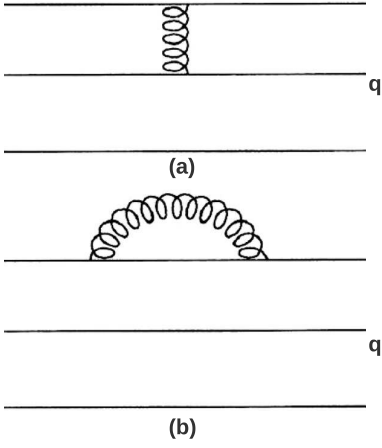}
\caption{\label{figgluon}One gluon exchange contributions to the baryon energy.}
\end{figure}

Finally, taking into account the specific quark flavor and spin configurations
in the ground state baryons and using the relations 
$\langle\sum_a(\lambda_i^a)^2\rangle =16/3$ and 
$\langle\sum_a(\lambda_i^a\lambda_j^a)\rangle_{i\ne j}=-8/3$ for
baryons, one can write the energy correction due to
color electric contribution, as
\begin{equation}
(\Delta E_N)_g^E={\alpha_c}(b_{uu}I_{uu}^E+b_{us}I_{us}^E+b_{ss}I_{ss}^E) \ ,   
\label{enge}
\end{equation}
and due to color magnetic contributions, as
\begin{equation}
(\Delta E_N)_g^M={\alpha_c}(a_{uu}I_{uu}^M+a_{us}I_{us}^M+a_{ss}I_{ss}^M) \ ,  
\label{engm}
\end{equation}
where $a_{ij}$ and $b_{ij}$ are the numerical coefficients depending on each 
baryon. In figure \ref{figgluon}, we have shown the one gluon exchange 
among the quarks. The color electric contributions for the baryon masses 
vanishes when all the constituent quark masses in a baryon are equal, 
whereas it is non-zero otherwise. Therefore, we have $a_{uu}=-3$ and  
$a_{us}=a_{ss}=b_{uu}=b_{us}=b_{ss}=0$ for the nucleon case.
The quantities $ I_{ij}^{E,M} $ are given in the following equation
\begin{eqnarray}
I_{ij}^{E}=\frac{16}{3{\sqrt \pi}}\frac{1}{R_{ij}}\Bigl[1-
\frac{\alpha_i+\alpha_j}{R_{ij}^2}+\frac{3\alpha_i\alpha_j}{R_{ij}^4}
\Bigl]
\nonumber\\
I_{ij}^{M}=\frac{256}{9{\sqrt \pi}}\frac{1}{R_{ij}^3}\frac{1}{(3\epsilon_i^{'}
+m_{i}^{'})}\frac{1}{(3\epsilon_j^{'}+m_{j}^{'})} \ ,
\end{eqnarray}
where 
\begin{eqnarray}
R_{ij}^{2}&=&3\Bigl[\frac{1}{({\epsilon_i^{'}}^2-{m_i^{'}}^2)}+
\frac{1}{({\epsilon_j^{'}}^2-{m_j^{'}}^2)}\Bigl]
\nonumber\\
\alpha_i&=&\frac{1}{ (\epsilon_i^{'}+m_i^{'})(3\epsilon_i^{'}+m_{i}^{'})} \ .
\end{eqnarray} 
In the calculation we have taken $\alpha_c= 0.58$ as the strong coupling
constant in QCD at the nucleon scale \cite{barik}. The color electric 
contribution is zero here, and the gluonic corrections to the mass of the 
nucleon are due to color magnetic contributions only. 

Finally treating all these corrections independently, one can obtain the 
physical mass of the nucleon as
\begin{equation}
M_N\equiv E_N=E_N^0-\epsilon_{cm}+\delta M_N^\pi+(\Delta E_N)^E_g+
(\Delta E_N)^M_g
\label{mass}
\end{equation}
where $\epsilon_{cm}$ is the energy associated with the spurious center
of mass correction, $(\Delta E_N)^E_g+(\Delta E_N)^M_g$ is the color electric 
and magnetic interaction energies arising out of the one-gluon exchange  
process and $\delta M^\pi_N$ is the pionic self-energy of the nucleon due
to pion coupling to the non-strange quarks. In the above $M_N$ is the mass 
of the nucleon at zero density. In the next section, we will calculate 
the effective mass $M_N^{*}$ in the medium using the above equation 
(\ref{mass}) where additional quark coupling to the mesons would be 
introduced in a mean field approximation.

\section{Equation of state for nuclear matter}

The Dirac equation (\ref{eq:3}) for  individual quarks in the medium is now
given by
\begin{equation}
[\gamma^0~(\epsilon_q-g_\omega^q\omega_0)-{\vec \gamma}.{\vec p}-
(m_q-g_\sigma^q\sigma)-U(r)]\psi_q(\vec r)=0 \ ,
\end{equation}
where $g_\sigma^q$ and $g_\omega^q$ are the quark couplings to  
the $\sigma$ and $\omega$ mesons.
In the mean field approximation, the meson fields are treated by their
expectation values,
\begin{equation}
\sigma\rightarrow\langle\sigma\rangle\equiv\sigma_0~~~~{\mbox and}~~~~
\omega_\mu\rightarrow\langle\omega_\mu\rangle\equiv\delta_{\mu 0}\omega_0 \ .
\end{equation}
We can now redefine equation (\ref{eq:9}) in medium as
\begin{equation}
\epsilon^{\prime}_q= (\epsilon_q^*-V_0/2)~~~ 
\mbox{and}~~~ m^{\prime}_q=(m_q^*+V_0/2),
\label{eprim}
\end{equation}
where the effective quark energy, 
$\epsilon_q^*=\epsilon_q- g_\omega^q\omega_0$ and effective quark mass, 
$m_q^*=m_q-g_\sigma^q\sigma_0$.
Substituting this in equation (\ref{eprim}), the effective mass of the 
nucleon at finite densities can be calculated from equation (\ref{mass})
\begin{equation}
M_N^*=E_N^* \ .
\end{equation}
The baryon density $\rho_B$, the total energy density and pressure at a 
particular baryon density for the symmetric nuclear matter are given in the 
usual form as:
\begin{equation}
\rho_B= \frac{\gamma}{(2\pi)^3}\sum_{N=p,n}\int_0^{k_f^N} d^3 k
=\frac{2 k_f^3}{3\pi^2} \equiv \rho_p+\rho_n \ .
\end{equation}

\begin{eqnarray}
{\cal E}&=&\frac{1}{2}m_\sigma^2 \sigma_0^2-\frac{1}{2}m_\omega^2 \omega^2_0
+g_\omega\omega_0\rho_B\nonumber\\ &+&
\frac{\gamma}{(2\pi)^3}\sum_{N=p,n}\int ^{k_f^N} d^3 k \sqrt{k^2+{M_N^*}^2},\\
P&=&-~\frac{1}{2}m_\sigma^2 \sigma_0^2+\frac{1}{2}m_\omega^2 \omega^2_0\nonumber\\
&+&\frac{\gamma}{3(2\pi)^3}\sum_{N=p,n}\int ^{k_f^N} \frac{k^2~ d^3 k}
{\sqrt{k^2+{M_N^*}^2}},
\end{eqnarray}
where $\gamma=2$ is the spin degeneracy factor for nuclear matter 
and $g_\omega=3g_\omega^q$ is the omega-nucleon coupling. 

The vector mean-field $\omega_0$ is determined through
\begin{equation}
\omega_0=\frac{g_\omega \rho_B}{m_\omega^2},
\label{omg}
\end{equation}
Finally, the scalar mean-field $\sigma_0$ is fixed by
\begin{equation}
\frac{\partial {\cal E}}{\partial \sigma_0}=0.
\label{sig}
\end{equation}
The scalar and vector couplings $g_\sigma^q$ and $g_\omega$
are fitted to the saturation density and binding energy for nuclear
matter. For a given baryon density, $\omega_0$ and $\sigma_0$ are
calculated from the equation (\ref{omg}) and (\ref{sig}) respectively.

The compressibility modulus $K$
is given by the standard relation:
\begin{eqnarray}
K & = & 9 \rho_B^2 \frac{\partial^2({\cal E}/\rho_B)}{ \partial\rho_B^2},
\label{e31}
\end{eqnarray}
which measures the stiffness of nuclear matter at the saturation point.

\section{Results and Discussion}

Starting with this simple composite model of nucleon in free space;
we wish to study several mean field properties of the nuclear matter, 
where the basic (N-N) interaction is realised through quark couplings to sigma 
($\sigma$) and omega  ($\omega$) mesons. We would also like to investigate the 
variations of these nuclear matter properties with the quark masses and their 
nuclear density dependance. 

\subsection{Free nucleon properties}
Apart from the bulk properties like binding energy and the compressibility, 
we would like to address few other properties of the nucleon in nuclear matter 
such as nucleon mass $M_N$, charge radius $<r^2>_N^{1/2}$, axial vector coupling
constant $g_A$, pion-nucleon coupling constant $g_{NN\pi}$,
and nucleon sigma term $\Sigma_N$. 

The expressions for 
$<r^2>_N^{1/2}$, $g_A$, $g_{NN\pi}$ in free space follows in  the 
present model according to Ref \cite{barik} as:
\begin{equation}
\langle r_N^2\rangle_{\rm without~ cm }=\frac{3}{2} \frac{11\epsilon_q^{\prime} 
+ m_q^{\prime}}{(3\epsilon_q^{\prime}+m_q^{\prime})(\epsilon_q^{\prime 2}-
m_q^{\prime 2})} \ ,
\end{equation}
and with cm correction
\begin{eqnarray}
\langle r_N^2\rangle &=&\langle B|{\frac{1}{3}}\sum_{r=1}^3({\vec r_i}-
{\vec R_{cm}})^2|B\rangle \nonumber\\
&=&\frac{11\epsilon_q^{\prime} +m_q^{\prime}}{(3\epsilon_q^{\prime}
+m_q^{\prime}) (\epsilon_q^{\prime 2}-m_q^{\prime 2})} \ .  
\end{eqnarray}
The axial-vector coupling constant
$g_A$ can also be obtained as \cite{barik}
\begin{equation}
g_A(n\rightarrow p)= \frac{5}{9}\frac{(5\epsilon_u^{\prime}
+7m_u^{\prime})} {(3\epsilon_u^{\prime}+m_u^{\prime})},
\end{equation}
without considering center of mass corrections. Another quantity of interest
is the quark-pion coupling constant $G_{qq\pi}$. Using the familiar Goldberger-Treiman
relation, we have,
\begin{equation}
\frac{G_{qq\pi}}{2M_q} = \frac{1}{2f_{\pi}}\times\frac{3}{5}g_A.
\end{equation}
Here $M_q$ is the constituent quark mass $M_N/3$. The pseudoscalar
pion-nucleon coupling constant $g_{NN\pi}$ which
is obtained from \cite{barik} at $q^2=m_{\pi}^2$
\begin{equation}
G_{NN\pi}(q^2)=\frac{M_N}{f_{\pi}}g_A u(q) \ ,
\end{equation}
where the axial-vector nucleon form-factor is
\begin{equation}
u(q)=\Big[1-\frac{3}{2} \frac{q^2}{{\lambda}_q(5\epsilon_q^{\prime}+
7m_q^{\prime})}\Big]e^{-q^2r_0^2/4} \ .
\end{equation}
The medium dependence of $g_A$, $G_{qq\pi}$ and $g_{NN\pi}$ will be discussed 
later.

We wish to study several mean field properties of our
composite model of the nucleon by fixing first the free-space
nucleon properties. Here, the quark mass $m_q$ is kept as a free parameter.
There are two unknown potential parameters $(a, V_0)$. These are obtained by 
fitting the nucleon mass $M_N=939$ MeV and charge radius of the proton
$\langle r_N\rangle=0.87$ fm in free space.
We point out here, that in the present 
model chiral symmetry is explicitly broken since the Lagrangian 
${\cal L}^0_q(x)$ is chirally odd with the explicit term 
$G(r) \bar\psi_q(x)\psi_q(x)$ where $G(r)=m_q+V(r)/2$. 
In view of PCAC, $m_q$ in the 
Lagrangian density is usually expected to be the curent mass. In the bag 
model picture, the quark mass are also taken in the current mass level.
Therefore, we investigate the variations of free space nucleon properties
vis-a-vis the saturation properties of nuclear matter with the light quark mass 
$m_q$ taken within a moderately low values $m_q=40$ and 50 MeV. 
However, if we consider $m_q$ to be an otherwise free parameter; we can also 
consider here $m_q$ = 300 MeV at the constituent mass range. With such 
choices of $m_q$ values we fit our basic inputs corresponding to the 
free-space nucleon properties together with the saturation properties 
of symmetric nuclear matter to determine the model parameters ($a$, $V_0$).

The parameters at zero
density and the energy contributions for various corrections to the nucleon
mass including the axial vector coupling constant $g_A$,  are given in
Table \ref{table1}.
\begin{table*}[t]
\centering
\renewcommand{\arraystretch}{1.4}
\setlength\tabcolsep{3pt}
\begin{tabular}{|c|c|c|c|c|c|c|c|c|c|}
\hline
$m_q$& $V_0$& $a$& $\epsilon_q$& $\epsilon_{c.m}$& $\delta M_N^\pi$ &
$(\delta E_B)_g$& $g_A$&$g_{NN\pi}$\\
(MeV) & (MeV) & (fm$^{-3}$) & (MeV) & (MeV) & (MeV) & (MeV) & &\\
\hline
40      & 100.187229 & 0.892380 & 483.516 & 373.636 &
-63.018  &-74.894&1.1179 &10.19\\
\hline
50      & 96.287247 & 0.870341 & 482.483 & 369.668 &
-65.749  &-73.302&1.1334 &10.34\\
\hline
300     & -62.257187 & 0.534296 & 458.455 & 283.578 &
-109.689  &-43.099&1.3844 &12.68\\
\hline
\end{tabular}
\caption{\label {table1} Bare set parameters and energy corrections.}
\end{table*}
Taking gluonic and pionic corrections together with that due to center of mass
correction, baryon mass spectra in vacuum had been already reproduced in 
similar potential model \cite{bdd}. The contributions due to pionic 
and gluonic corrections for
octet baryons and $\Delta$ have been described in detail \cite{bdd}. 

In the present work, since our focus is on the study of the nuclear matter,
we did not make fine tunning to reproduce the mass of the baryon spectra. The
gluon contribution to the nucleon in the present calculation comes out to be
about $-75$ MeV. The pionic correction to $\Delta$ is $-\frac{99}{171}\delta
M^\pi_N$ and the gluonic correction to $\Delta$ as realized by putting
$a_{uu}=3$ and $a_{us}=a_{ss}=b_{uu}=b_{us}=b_{ss}=0$ in equation 
(\ref{enge}) and (\ref{engm}), we get $M_\Delta=1115.32$ MeV.

\subsection{Nucleons in medium and equation of state}
We next fix the couplings, $g_\sigma^q$ and $g_\omega$, by fitting saturation
properties of the nuclear matter.
We take the standard values for the meson masses, namely $m_\sigma=550$ MeV
and $m_\omega=783$ MeV.
The quark-meson coupling constants $g_\sigma^q$ , $g_\omega=3g_\omega^q$
are fitted self-consistently to obtain the correct saturation properties
of nuclear matter binding
energy, $B.E \equiv{\cal E}/\rho_B - M_N 
= -15.7$~MeV and pressure, $P=0$ at $\rho_B=\rho_0=0.15$~fm$^{-3}$.
These parameters are given in Table \ref{table2}.
Due to the additional quark-meson coupling in the nuclear matter 
representing (N-N) interaction, since effective quark mass becomes 
$m_q^*$ = ($m_q$ - $g_\sigma^q$$\sigma_0$). 
The compressibility comes around
223 MeV for $m_q=40$ MeV, 224 MeV for $m_q=50$ MeV and 259 MeV for $m_q=300$ MeV at
nuclear matter density which is usually taken to be about 200-300 MeV.
The mean field values of $\sigma_0$ and $\omega_0$, compressibility,
effective mass, the energy contributions from
different corrections, $g_A$ and $g_{NN\pi}$ at the saturation point are
provided in Table \ref{table3}.
\begin{table}[t]
\centering
\renewcommand{\arraystretch}{1.4}
\setlength\tabcolsep{3pt}
\begin{tabular}{|c|c|c|}
\hline
$m_q$ (MeV)& $g^q_\sigma$& $g_\omega$\\
\hline
40     &5.46761  &3.96975  \\
\hline
50      &5.28816 & 4.30828 \\
\hline
300   &4.07565  &9.09078  \\
\hline
\end{tabular}
\caption{\label{table2}Parameters for nuclear matter. They are determined
from the binding energy per nucleon, $B.E \equiv{\cal E}/\rho_B - M_N 
= -15.7$~MeV and pressure, $P=0$ at saturation density $\rho_B=\rho_0=0.15$~fm$^{-3}$.}
\end{table}

\begin{table*}[t]
\centering
\renewcommand{\arraystretch}{1.4}
\setlength\tabcolsep{2pt}
\begin{center}
\begin{tabular}{|c|c|c|c|c|c|c|c|c|c|c|c|c|}
\hline
$m_q$&$\sigma_0$&$\omega_0$&$M_N^*/M_N$&K & $\langle r_{N}\rangle$&$\epsilon_q$&
$\epsilon_{c.m}$& $\delta M_N^\pi$ & $(\delta E_B)_g$&$g_A$&$g_{NN\pi}$\\
(MeV) & (MeV) & (MeV) &  & (MeV) & (fm) & (MeV) & (MeV) & (MeV) & (MeV) &  & \\
\hline
40 &15.07&7.46&0.90 & 222.48 & 0.94784 & 447.884 & 376.091 &-28.909&
-82.861&0.9448&7.64\\
\hline
50 &15.74&8.09&0.90 &223.81 & 0.94794 & 445.232 & 373.260 &-30.545&
-81.603  &0.9624&7.74\\
\hline
300&26.93&17.08&0.77 & 258.913 & 0.94902 & 382.049 & 303.502 &-65.282&
-53.855&1.2629&8.73\\
\hline
\end{tabular}
\vspace{0.5cm}
\caption{\label{table3}Properties for nuclear matter at saturation density. }
\end{center}
\end{table*}
The binding energy per nucleon for nuclear matter as a function of nucleon
density $\rho_N$ corresponding to each of the choices of $m_q$ values have been 
calculated. Therefore in Fig. 2, we plot this result for $m_q$ = 40 MeV, 50 MeV
and 300 MeV to compare our result with that of NL3 \cite{nl3} and
QMC \cite{qmcparm}. In the same figure, the equation of state for neutron matter 
in the present model with $m_q$ = 40 MeV, 50 MeV and 300 MeV has also been 
depicted.
\begin{figure*}
\begin{center}
\includegraphics[width=7.cm,angle=0]{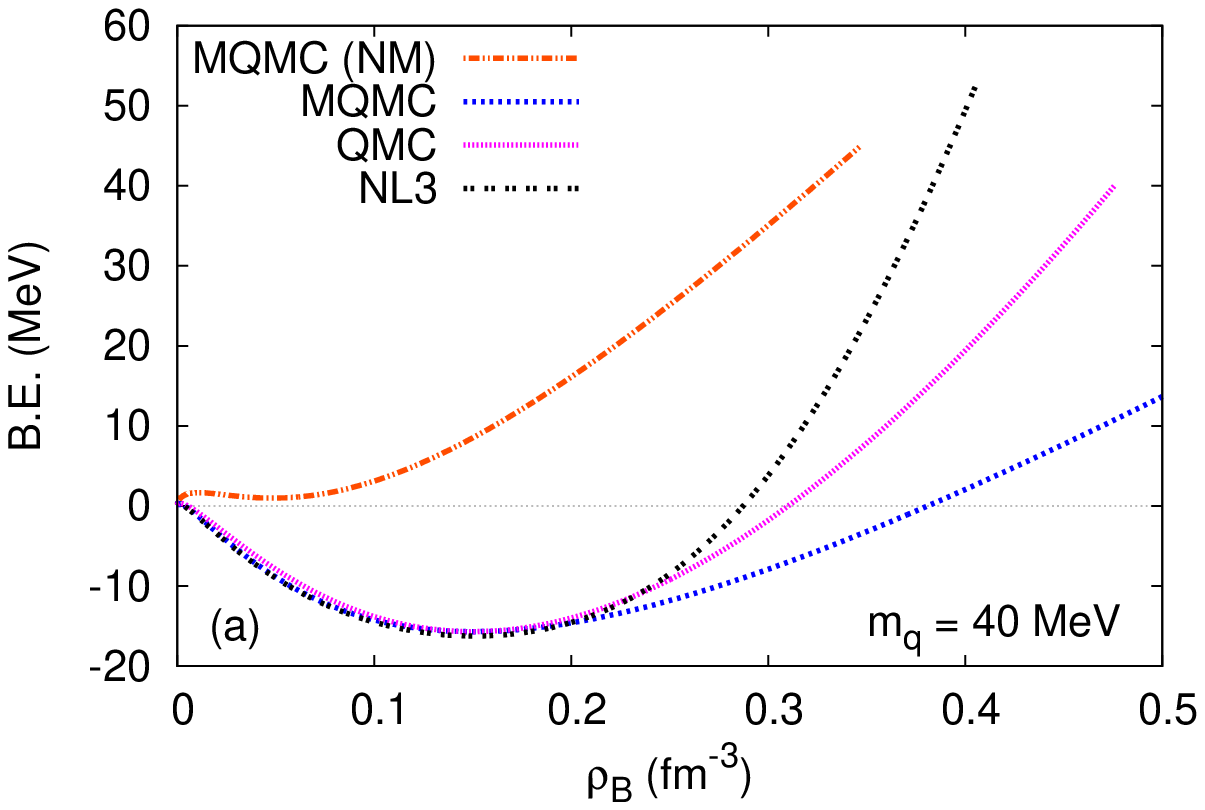}
\includegraphics[width=7.cm,angle=0]{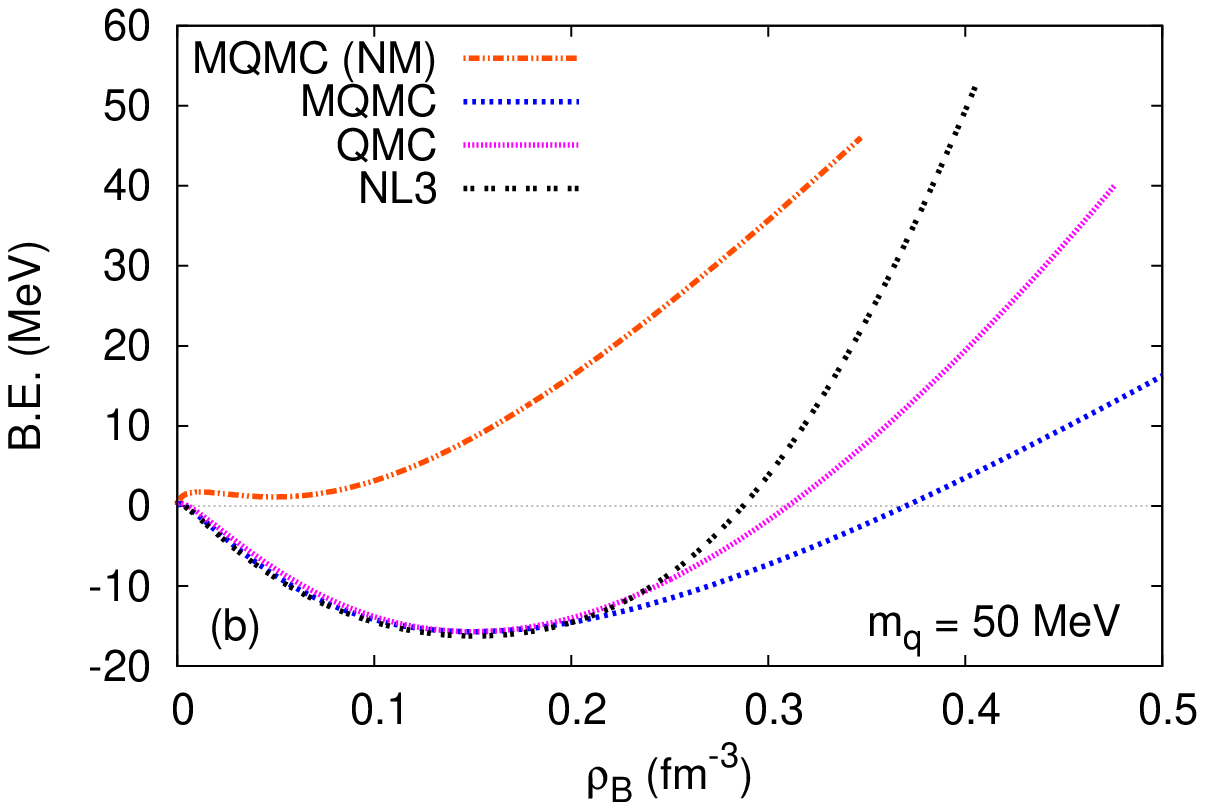}
\includegraphics[width=7.cm,angle=0]{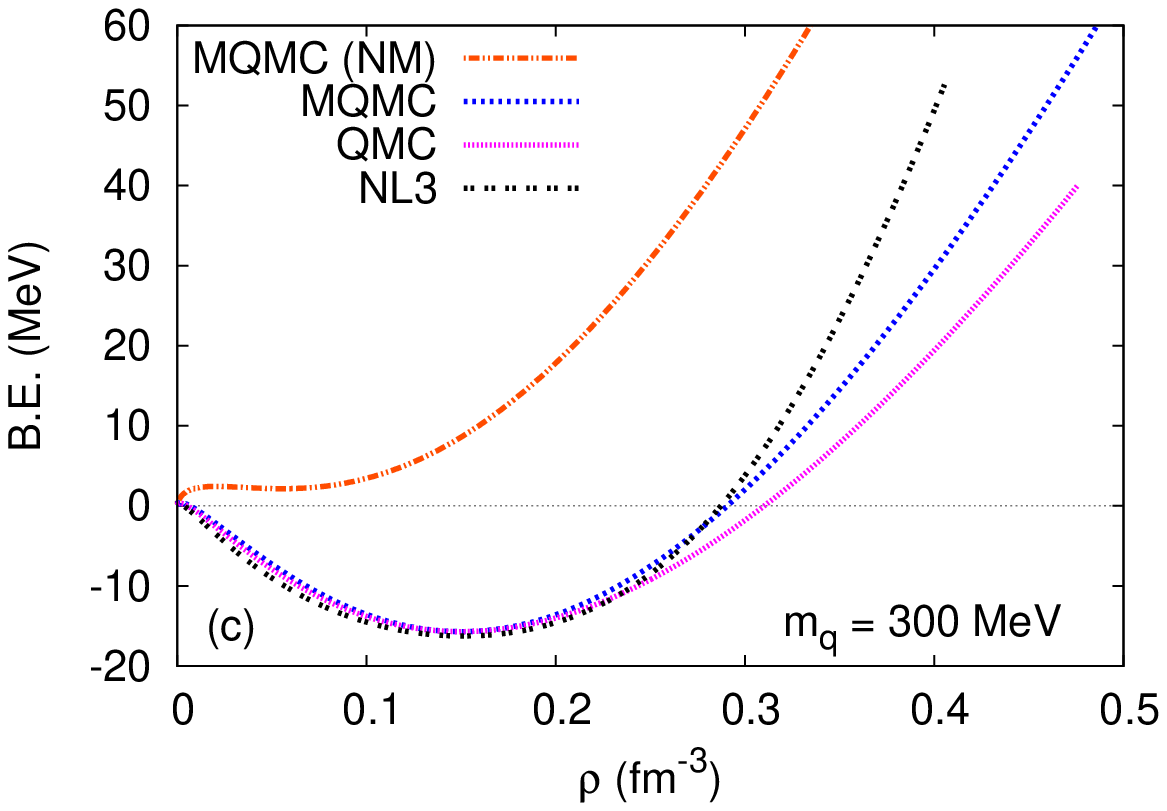}
\end{center}
\caption{Nuclear matter binding energy as a function of density for
different quark mass.}
\label{fig1}
\end{figure*}

\begin{figure}
\includegraphics[width=8.cm,angle=0]{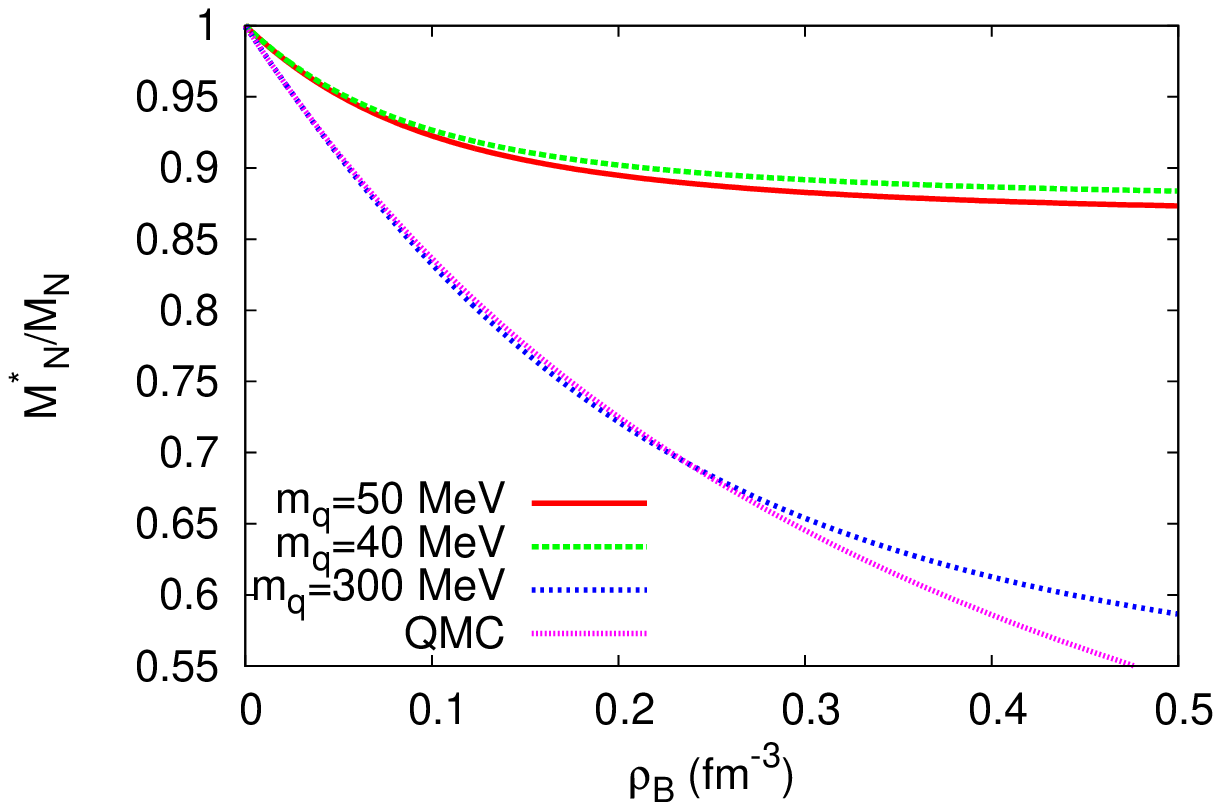}
\caption{\label{fig2}Effective mass versus density with quark mass 
$m_q=40$ MeV, $m_q=50$ MeV and $m_q=300$ MeV.}
\end{figure}
Figure \ref{fig2} shows the effective nucleon mass, $M^*/M$, as a 
function of baryon density for quark mass $m_q=40$ MeV, 50 MeV and 300 MeV.
This result is compared with that obtained in QMC \cite{qmcparm}.
In all cases, the effective mass decreases as the baryon 
density increases and then saturates at high baryon densities.
It may be noted here that the effective nucleon mass of the Walecka 
model\cite{qhd} at saturation density is about 540 MeV, a value considered 
to be extremely small. A value of about 700 to 750 MeV is usually obtained in
nonrelativistic calculations which is  considered to be more consistent with 
the observed value of the density of states near the Fermi surface. 
In QMC, $M_N^*$ was found to be of the order of 723 MeV. However,
in the present analysis the effective mass $M_N^*$ comes out to be
855 MeV with $m_q=40$ MeV, 850 MeV with $m_q=50$ MeV and 723 MeV with $m_q=300$ 
MeV. Saito and Thomas \cite{ST} model obtained $M_N^*=839-856$ MeV with the 
bag radius varying in the range of 0.6 fm to 1 fm.

We next calculate the spin-orbit potential strength, $V_{so}$, 
following Ref.  \cite{batista} and found that $V_{so}= 0.75$ MeV for $m_q=40$
MeV, $V_{so}= 0.83$ MeV for $m_q=50$ MeV and $V_{so}= 3.41$ for $m_q=300$ MeV.
It rises smoothly with increasing quark mass. Phenomenological value of the
spin-orbit strength are in the range from 5 to 7 MeV.

\begin{figure}
\includegraphics[width=8.cm,angle=0]{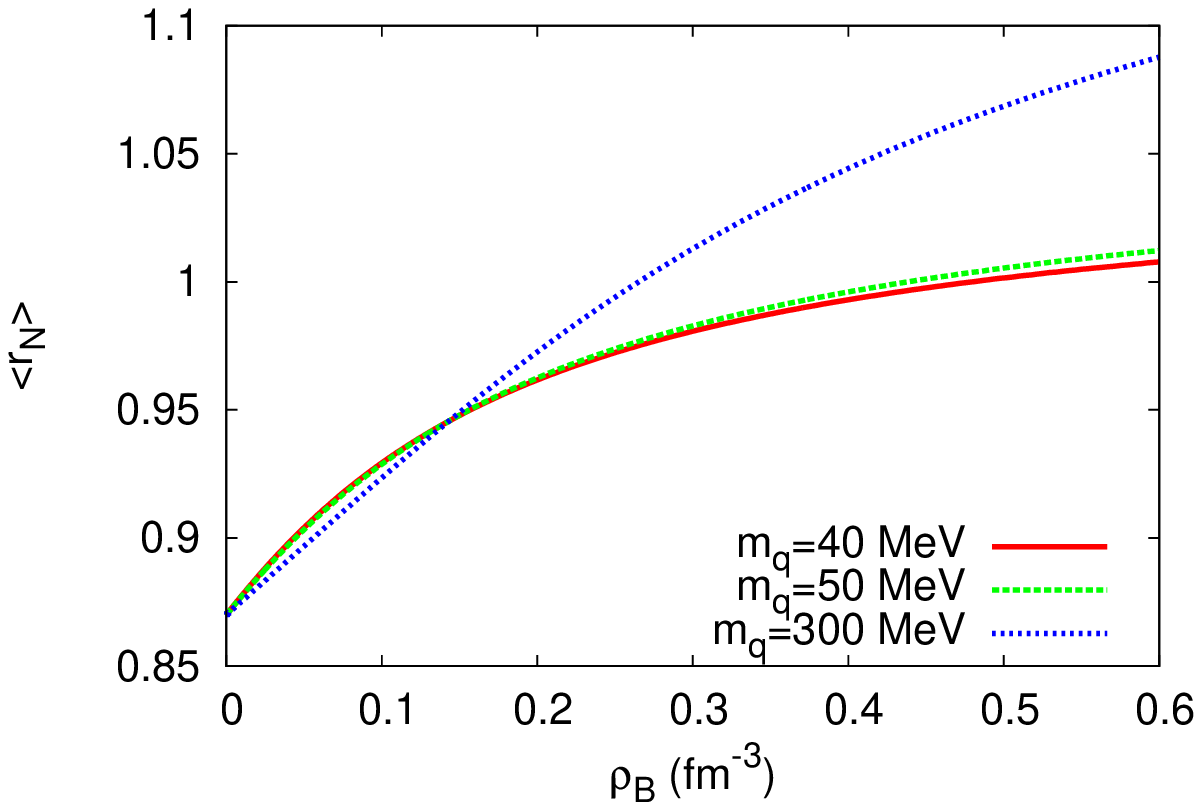}
\caption{\label{fig3}Charge radius versus density with quark mass 
$m_q=40$ MeV, $m_q=50$ MeV and $m_q=300$ MeV.}
\end{figure}
The variations of the root mean square
nucleon radius, $r_N$, are shown in Figure \ref{fig3} with baryon density 
for quark mass $ m_q=40$ MeV, $m_q=50$ MeV and $m_q=300$ MeV. The nucleon radius
increases with the baryon density and is approximately 0.95 fm at the
saturation density. The rate of increase 
tends to be larger for larger values of the quark mass.

\begin{figure}
\includegraphics[width=8.cm,angle=0]{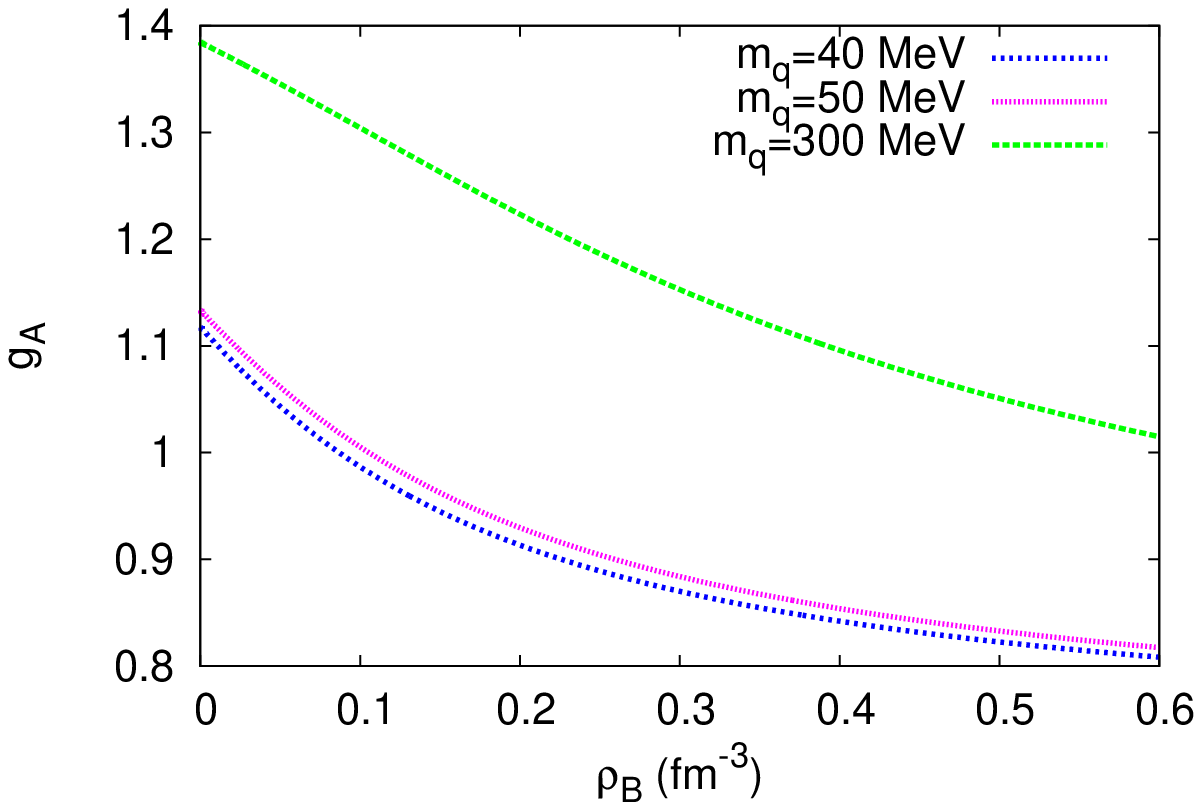}
\caption{\label{fig4}$g_A$ versus density with quark mass $m_q=50$ MeV,
$m_q=50$ MeV and $m_q=300$ MeV.}
\end{figure}
In  Figure \ref{fig4}, the variation of axial vector coupling 
constant  $g_A$ as a function of baryon density for quark mass 
$m_q=40$ MeV, $m_q=50$ MeV and $m_q=300$ MeV are shown. At bare level, 
the $g_A=1.118$ for $m_q=40$ MeV, $g_A=1.133$ for $m_q=50 $ MeV and $g_A=1.384$ 
for $m_q=300 $ MeV which qualitatively agrees with
the experimental value  $g_A/g_V = 1.27590_{-0.00445}^{+0.00409}$ \cite{liu}.
$g_A$ is observed to decrease with increase in the
density. At saturation density, $g_A=0.945$ for $m_q=40$ MeV, $g_A=0.962$
for $m_q=50$ MeV and $g_A=1.263$ for $m_q=300$ MeV. Since,
our MQMC model is a relativistic model, the attractive scalar potential 
decreases the quark mass. Thus the lower component of the wave function 
is enhanced and hence it makes $g_A$ decrease with density. This is similar
to the observations made in \cite{saitoprc}.
\begin{figure*}
\includegraphics[width=7.cm,angle=0]{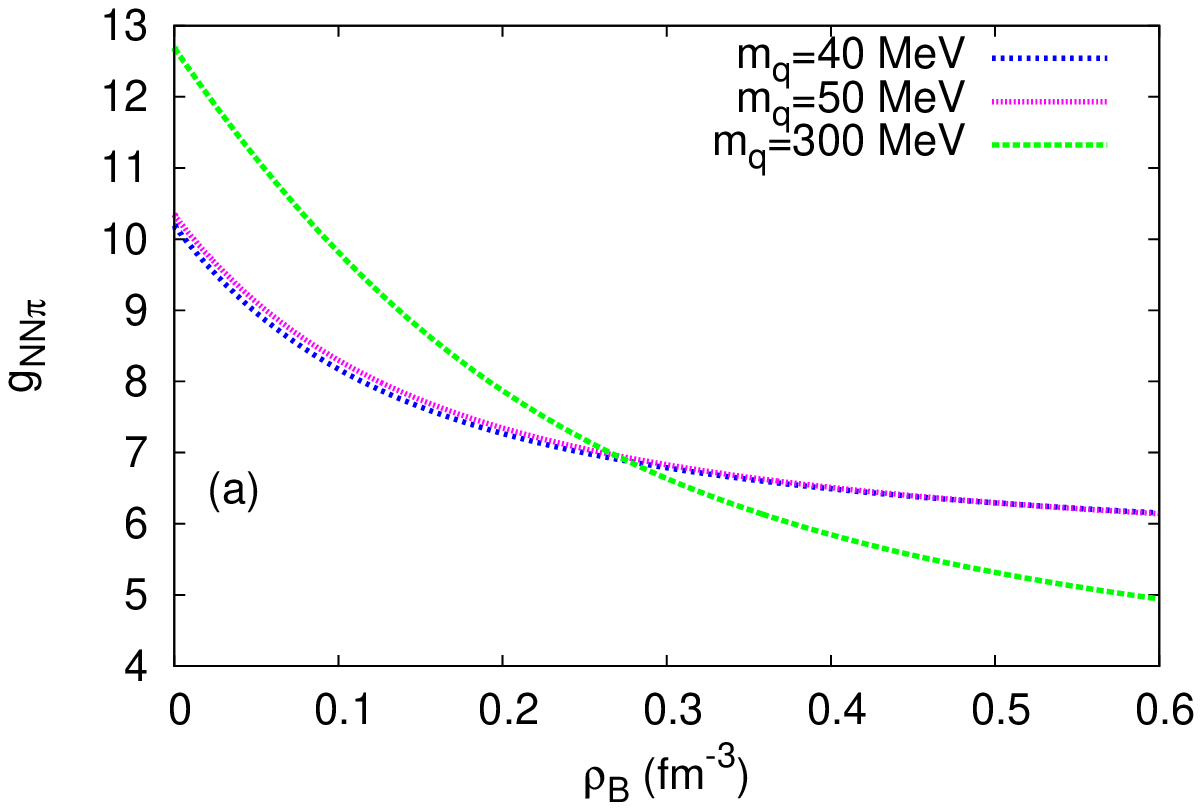}
\includegraphics[width=7.cm,angle=0]{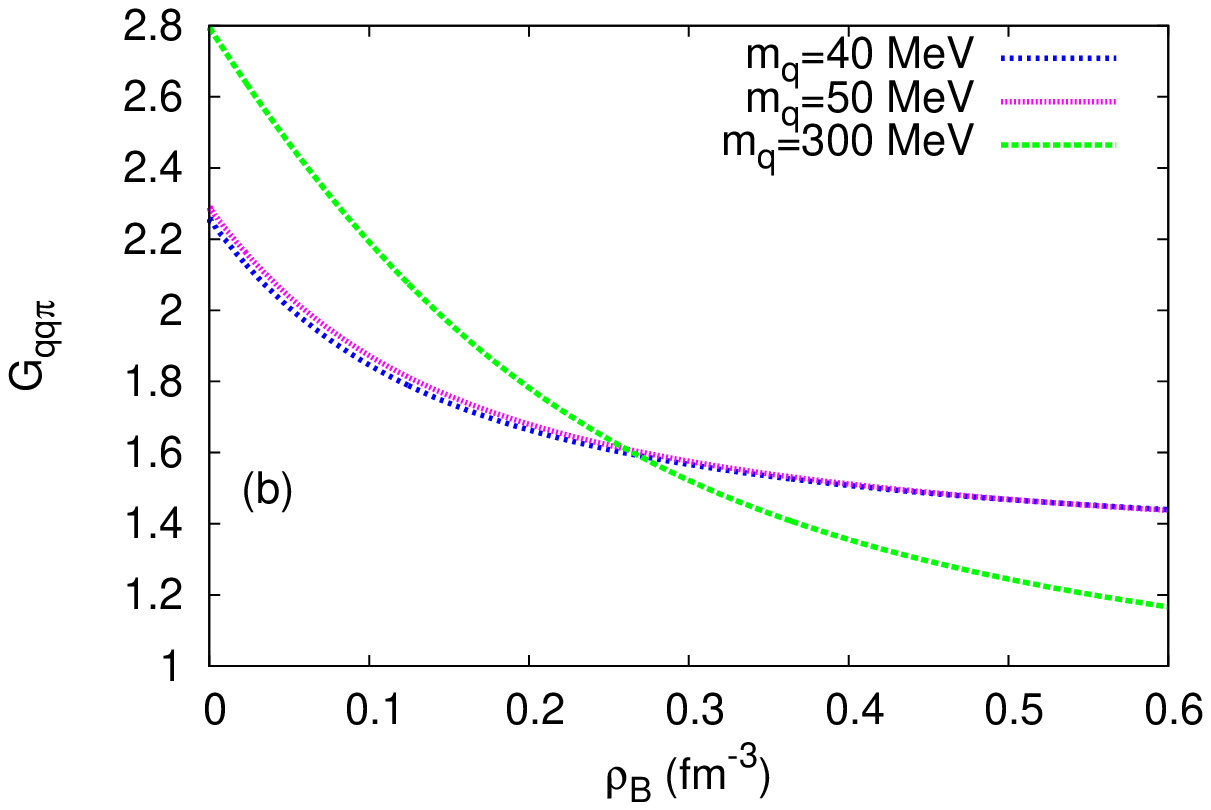}
\caption{\label{fig5-6}(a) $g_{NN\pi}$ versus baryon density with quark mass 
$m_q=40$ MeV, $m_q=50$ MeV and $m_q=300$ MeV.
(b) $g_{qq\pi}$ versus baryon density with quark mass 
$m_q=40$ MeV, $m_q=50$ MeV and $m_q=300$ MeV.}
\end{figure*}
The nucleon-pion coupling constant $g_{NN\pi}$ and quark
pion coupling constant $g_{qq\pi}$ with $m_q=40$, 50 MeV  and 300 MeV as a function
of density are plotted in Figure \ref{fig5-6}(a) and \ref{fig5-6}(b) respectively. 
It is observed that both $g_{NN\pi}$ and $g_{qq\pi}$ 
decrease by increasing the density. This is due to the similar trend found
in $g_A$ since they are related through Goldberger-Treiman relation.

\begin{figure}
\includegraphics[width=7.cm,angle=0]{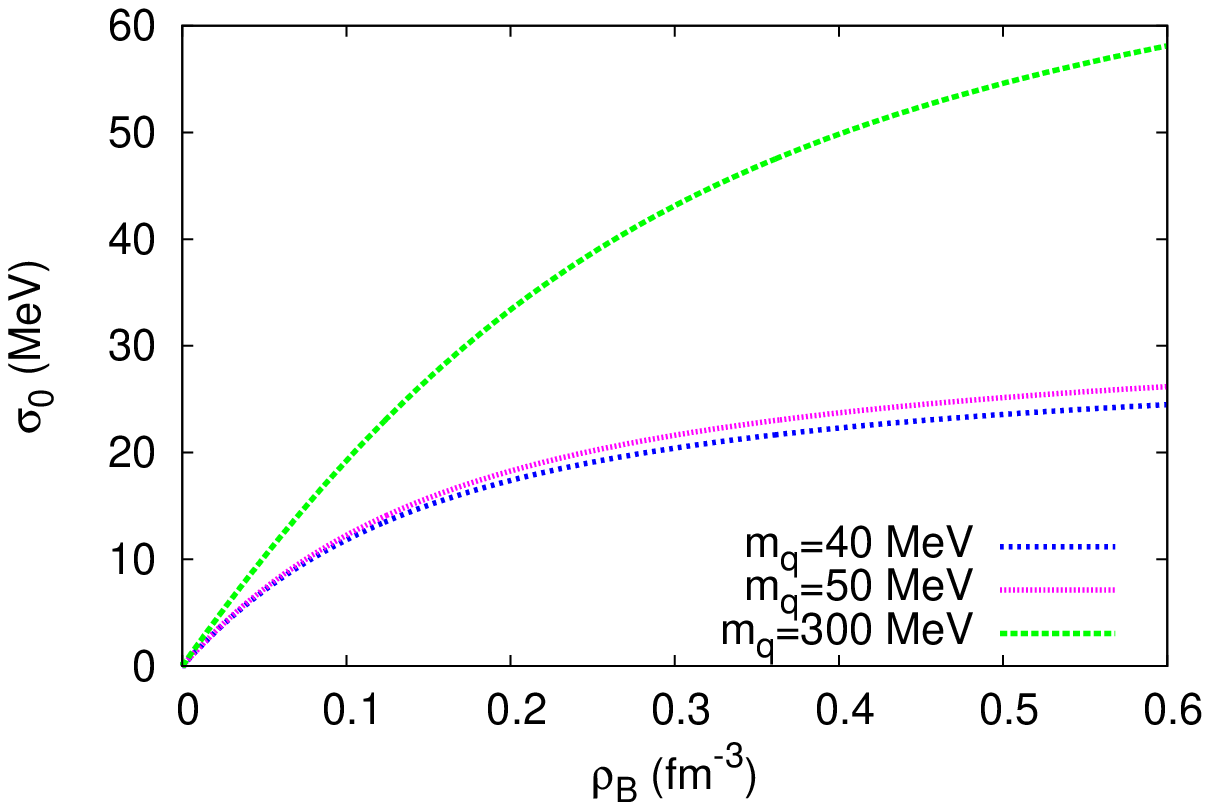}
\caption{\label{fig8} $\sigma_0$ versus baryon density with quark mass 
$m_q=40$ MeV, $m_q=50$ MeV and $m_q=50$ MeV.}
\end{figure}
We have calculated the scalar mean field $\sigma_0$ at various densities which
is plotted in \ref{fig8} for $m_q=40$ MeV, $m_q=50$ MeV $m_q=300$ MeV. At 
saturation density, we find $\sigma_0=15.44$ MeV for $m_q=40$ MeV, 
$\sigma_0=16.11$ MeV for $m_q=50$ MeV and $\sigma_0=26.93$ MeV for $m_q=300$ MeV.
It is quite interesting to note here that the effective mass 
of the quark that has entered in our calculation as $m_q^\prime=
m_q-g_\sigma^q\sigma_0+V_0/2$ comes out to be 7.697  MeV at
$m_q=40$ MeV and 14.9 MeV for $m_q=50$ MeV at saturation density.
Such low effective quark mass, which is of the order of up-down current 
quark masses is in quite commensurate consistency with PCAC requirement.

\begin{figure}
\includegraphics[width=8.cm,angle=0]{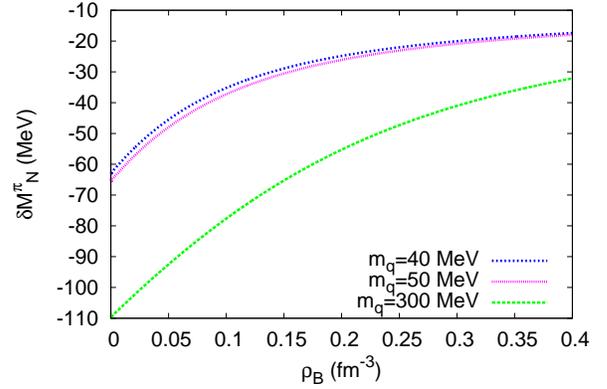}
\caption{\label{fig10}$\delta M^\pi_B$ versus baryon density with quark mass
$m_q=40$, $m_q=50$ MeV and $m_q=300$ MeV.}
\end{figure}
In Figure \ref{fig10}, the pionic corrections ${\delta M_N^{\pi}}$ to the mass
of the nucleon for quark masses 40 MeV, 50 MeV and 300 MeV are shown for
different baryon densities. It is found that ${\delta M_N^{\pi}}$ increases with
density and at saturation density
the values are -29 MeV, -30 MeV  and -65 MeV for the quark masses 40, 50 and 300
MeV respectively. Since with increase in density
the quark-pion coupling strength and the pseudoscalar nucleon-pion coupling
$g_{NN\pi}$ decreases, the pionic correction to the mass increases.

\begin{figure}
\includegraphics[width=8.cm,angle=0]{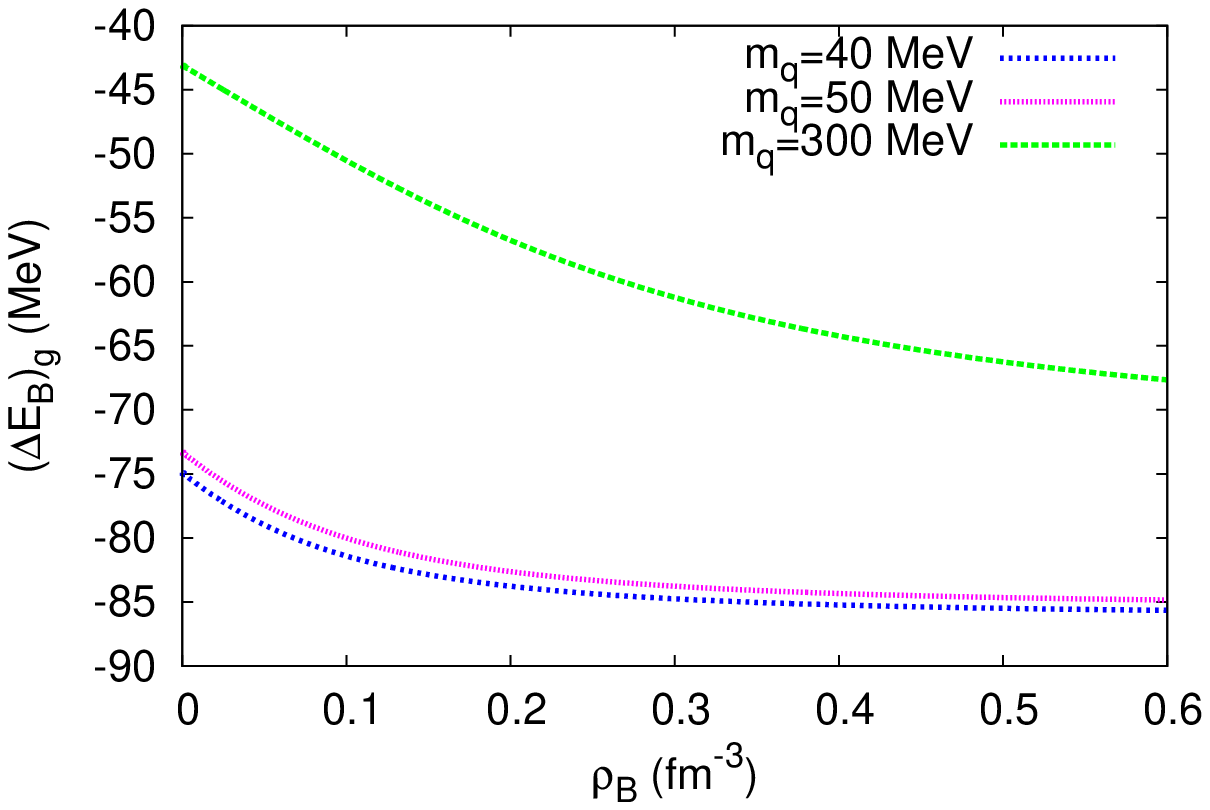}
\caption{\label{fig11}$(\Delta E_B)_g$ versus baryon density with quark mass
$m_q=40$, $m_q=50$ MeV and $m_q=50$ MeV.}
\end{figure}
In Figure \ref{fig11}, it is observed that the gluonic corrections to the mass
of the nucleon decreases by increasing the baryon density, which is expected, 
because the average quark distances increase as the nucleon swells.
In the same figure,  the gluonic correction for the quark 
masses 40, 50 and 300 MeV are compared. The rate of fall appears to be the 
same for different quark masses.

\subsection{Nucleon and nuclear matter sigma terms}
\begin{figure}[h]
\includegraphics[width=8.cm,angle=0]{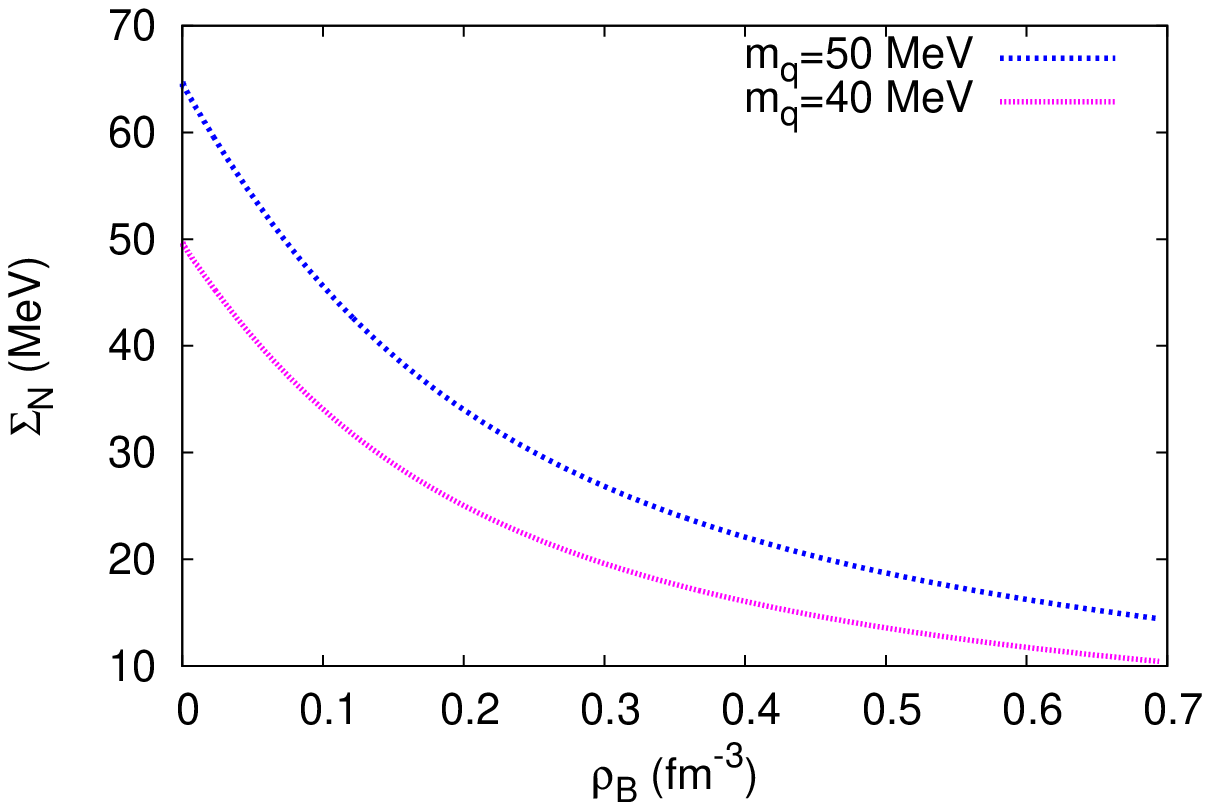}
\caption{\label{fig12}Nucleon sigma term versus baryon density with quark mass
$m_q=40$ and $m_q=50$ MeV.}
\end{figure}
We next proceed to calculate the nucleon sigma term, $\Sigma_N$
which is an important property for chiral symmetry.
The individual nucleon sigma term in the nuclear medium can be defined as 
(see \cite{cohen,gammal})
\begin{equation}
\Sigma_N=m_q\frac{\partial M_N}{\partial m_q} \ ,
\label{sig1}
\end{equation}
from the  Feynman-Hellman theorem. Note that, $M_N$ is identified with 
$M^*_N$ at finite density.
Alternatively, the nucleon sigma term $\Sigma_N$ can be related to the
quark condensates at low densities as \cite{cohen}
\begin{eqnarray}
2m_q[\langle \bar q q\rangle_{\rho_B}-\langle \bar q q\rangle_{vac}]&=&
\Sigma_N\rho_B+\cdots\nonumber\\
&=&m_q\frac{\partial \cal E}{\partial m_q}
\label{sig2}
\end{eqnarray}
where $\cal E$ is the energy density of nuclear matter, which is given as
\begin{equation}
{\cal E}=M_N^*\rho_B+\delta{\cal E}.
\end{equation}
\begin{figure}[h]
\includegraphics[width=7.cm,angle=0]{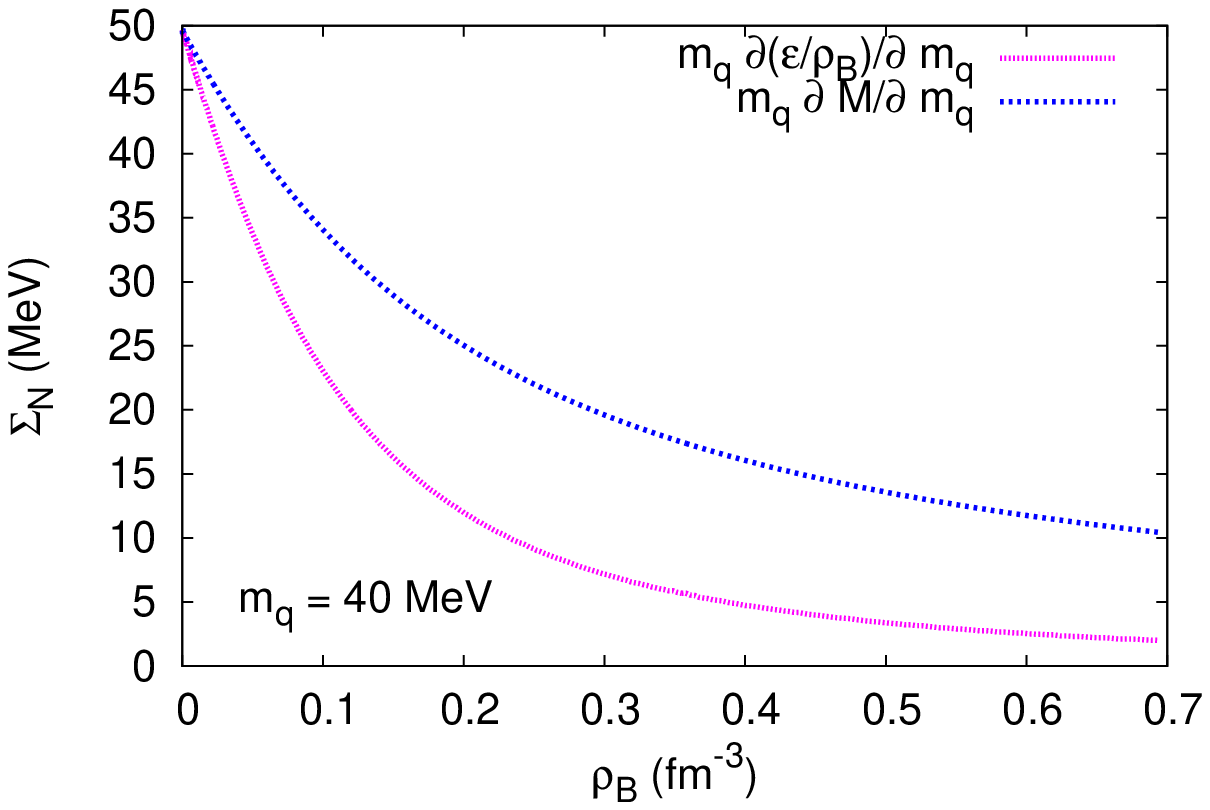}
\caption{\label{fig13}Nuclear and nucleon sigma terms versus baryon density 
with quark mass $m_q=40$ MeV.}
\end{figure}
Here $\delta{\cal E}$ in the calculation to the energy density from the 
nucleon kinetic energy plus the nucleon-nucleon interaction. $\delta{\cal E}$ 
is said to be small at low densities. Then, from equation (\ref{sig2}) one
can obtain the nuclear matter sigma term per nucleon as
\begin{equation}
\Sigma_{NM} =m_q\frac{\partial ({\cal E}/\rho_B)}{\partial m_q}
\label{sig3} \ ,
\end{equation}
which is distinct from the individual nucleon sigma term in nuclear matter, due
to the nucleon kinetic energy and the interaction among the nucleons in the 
nuclear medium. Now, we calculate $\Sigma_N$ using (\ref{sig1}) and 
$\Sigma_{NM}$(\ref{sig3}) at
various densities with the results shown in Figure \ref{fig12}
and Figure \ref{fig13} respectively for  $m_q=40$  and 50 MeV. 

The difference in the two results from (\ref{sig1}) and (\ref{sig3}) at 
various densities are plotted in Figure \ref{fig13}. At zero density we 
found $\Sigma_N^0=49.59$ MeV for $m_q=40$ MeV and $\Sigma_N^0=64.823$ MeV 
for $m_q=50$ MeV. The experimental value extracted from 
pion-nucleon scattering is 
$\Sigma_N \sim 45$ MeV \cite{gasser}. At saturation density, we find $\Sigma_N$ to be
$\Sigma_N^0=28.8$ MeV for $m_q=40$ MeV $\Sigma_N^0=38.9$ MeV for $m_q=50$ MeV.

\begin{figure}[h]
\includegraphics[width=7.cm,angle=0]{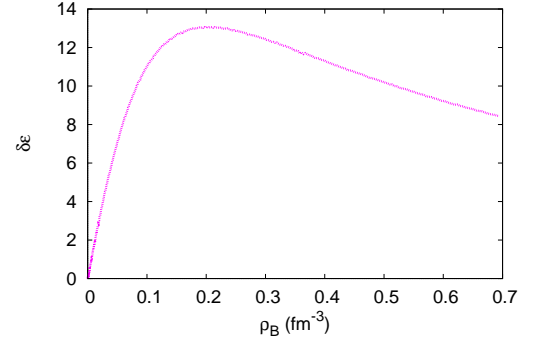}
\caption{\label{fig14}Variation of nuclear and nucleon sigma term versus 
baryon density with quark mass $m_q=40$ MeV.}
\end{figure}

The sensible quantity for the stability of nuclei is the relative variation 
of the binding energies with the quark mass, written below for nuclear matter:
\begin{eqnarray}
\frac{\delta BE}{BE}=\frac{\delta\left[{\cal E}/\rho_B-M_N\right]}
{{\cal E}/\rho_B-M_N}=
K_{NM}(\rho_B) \frac{\delta m_q}{m_q} , 
\label{vb}
\end{eqnarray}
which gives $K_{NM}=-1.02$ at nuclear saturation density. 
The general trend of $K_{NM}(\rho_B)$ with
the density, follows from the results presented in figure \ref{fig12}, 
it is negative and the magnitude
increases for larger densities, as we see in figure \ref{fig14}, which suggests
that compact objects could be more sensitive to variations in quark masses.  
We compare our result of $K_{NM}=$-1.02 at 
the saturation density with the values of the sensitivity $K_A$ for nuclei 
A=3-8 found in the range −1 to −1.5, computed in \cite{flambaum07} for 
different Argonne potential models, considering cases
including the Urbana model IX three-body force, and for thorium 
$K_{^{229}Th}=-1.45$\cite{flambaum09}. More recently, 
ref. \cite{carillo} computed for oxygen $K_{^{16}O}=-1.082$, and the 
sensitivity for other light nuclei with
a one boson exchange model.
It is to be noted here that in the present model the ratio of the quark
condensate in the leading order  
\begin{equation}
\frac{\langle \bar q q\rangle_{\rho_B}}{\langle \bar q q\rangle_{vac}}=
1-\frac{\Sigma_N\rho_B}{m_\pi^2f_\pi^2}\simeq\left\{\begin{array}{c}
0.80 \mbox{ for $m_q=40$ MeV}\\
0.74 \mbox{ for $m_q=50$ MeV}
\end{array}
\right. 
\end{equation}
are somewhat smaller than the results found in Ref \cite{gammal} with 
a Skyrme model of the nucleon and of the nuclear force. 
\par
Comparing our 
result with the ratio $M_N^*/M_N$ = 0.9 we find that the condensate 
ratio at saturation density (low density) essentially comes out as $\simeq$
$(M_N^*/M_N)^2$ as found by Saito and Thomas \cite{ST}. This result is
intermediate between the cubic dependence found by Brown and Rho \cite{brown}
and the linear dependence proposed by Cohen {\it et. al.} \cite{cohen}.
However, if we take the quark mass $m_q$ as 300 MeV, we also get the 
saturation property with reasonable agreement with the standard values
except for the nucleon sigma term. At $m_q$ = 300 MeV, the value of 
$\Sigma_N^0$ is much higher as compared to the experimental value.

Note that the quark mass at zero barionic density is
few times larger than the current up-down quark masses, which was necessary 
in order to approach the nucleon sigma term in the vacuum. Presumably, such 
value of the quark mass is parameterizing the complexity of the nucleon 
wave function beyond the valence state, which should contribute to the 
nucleon and nuclear matter matrix element of the $\bar q q$ operator.
However, we expect that the typical changes in the sigma term due to the 
nuclear environment will be kept in more refined description of the nucleon 
wave function. Noteworthy to mention that, although we have employed such 
simplified nucleon, we have obtained  results for 
the sensitivity of the nuclear binding energy comparable with the ones 
found in previous studies.

\section{Conclusion}

In the present paper we have studied the EOS for nuclear matter using
a modified-quark-meson-coupling-model (MQMC). The properties of nuclear matter were calculated 
relying in a self-consistent method starting with  a relativistic quark  model with chiral symmetry 
for independent nucleons. The nucleon in the nuclear medium is composed 
by the three independent 
relativistic quarks confined by an equal admixture  of scalar-vector harmonic potential in a background
of scalar and vector mean fields. We computed the corrections from the 
center of mass motion, pionic and gluonic exchanges within the nucleon 
to obtain its effective mass. The nucleon-nucleon interaction
in nuclear matter is then realized by the quark coupling to the scalar 
(sigma) and vector (omega) mesons through a mean field approximation.
Several basic characteristics of nuclear matter, such as the compressibility,
the nucleon effective mass and nuclear sigma term show better agreement 
with the experimental data than those obtained in a model with point-like 
nucleons. We have compared our results obtained with the quark-meson-coupling
model which is based on the bag model and with the ones obtained within the 
non-linear Walecka model. 

The sensitivity of the nuclear binding energy was computed giving 
$K_{NM}\simeq -1$ ($\delta B_E/B_E=K_{NM} \delta m_q/m_q$), 
and we found that the sensitivity 
rises with density, as the nuclear sigma term tends to vanish for 
large densities. The calculation of $K_{NM}$ receives sizable
effects from the nuclear interaction and kinetic energy, which 
decreases the sensitivity by almost a factor of 2 from the value computed
only considering the individual nucleon sigma term at the nuclear matter 
density. Finally, we have to mention that the model quark mass at zero barionic density is
few times larger than the current up-down quark masses, in order to approach the nucleon sigma term in 
the vacuum. This quark mass is effectively taken care of the complexity of the nucleon wave function 
beyond the valence state, assumed here, which of course should contribute to the mean value of the 
$\bar q q$ operator in the nucleon and nuclear matter states.
We expect that the typical changes in the sigma term due to the nuclear environment will be kept 
in more realistic descriptions of the nucleon, beyond the valence state, as our comparison  with other 
models show that the values of the sensitivity are quite compatible.  Further implications of this model 
for nuclear matter and compact stars, would be taken up in our future work.
\section*{ACKNOWLEDGMENTS} TF thanks Funda\c c\~ao de Amparo \`a Pesquisa do Estado de S\~ao
Paulo  (FAPESP) and to Conselho Nacional de Desenvolvimento Cient\'fico e Tecnol\'ogico (CNPq) of Brazil.

\end{document}